\documentclass{iau}
\usepackage{graphicx}

\newcommand{\be}{\begin{equation}}
\newcommand{\ee}{\end{equation}}
\newcommand{\ba}{\begin{eqnarray}}
\newcommand{\ea}{\end{eqnarray}}

\newcommand{\mpc}{\rm {h^{-1}Mpc }}
\newcommand{\ltsima}{$\; \buildrel < \over \sim \;$}
\newcommand{\lsim}{\lower.5ex\hbox{\ltsima}}
\newcommand{\gtsima}{$\; \buildrel > \over \sim \;$}
\newcommand{\gsim}{\lower.5ex\hbox{\gtsima}}

\def\gtrsim{\mathrel{\hbox{\rlap{\hbox{\lower4pt\hbox{$\sim$}}}\hbox{$>$}}}}

\def\lesssim{\mathrel{\hbox{\rlap{\hbox{\lower4pt\hbox{$\sim$}}}\hbox{$<$}}}}

\newcommand{\LCDM}{$\Lambda$CDM}

\title[Supervoid at the CMB Cold Spot] 
{Supervoid Origin of the Cold Spot in the Cosmic Microwave Background}

\author[Andr\'as Kov\'acs, Istv\'an Szapudi, Benjamin R. Granett et al.]
{
Andr\'as Kov\'acs$^{1,2}$, Istv\'an Szapudi$^{3}$, Benjamin R. Granett$^{4}$, Zsolt Frei$^{1,2}$, Joseph Silk$^5$,
Will Burgett$^{3}$, Shaun Cole$^{6}$, Peter W. Draper$^{6}$, Daniel J. Farrow$^{6}$, Nicholas Kaiser$^{3}$,
Eugene A. Magnier$^{3}$, Nigel Metcalfe$^{6}$, Jeffrey S. Morgan$^{3}$, Paul Price$^{7}$, John Tonry$^{3}$,  Richard Wainscoat$^{3}$
}

\affiliation{
$^{1}$ Institute of Physics, E\"otv\"os Lor\'and University, 1117 P\'azm\'any P\'eter s\'et\'any 1/A Budapest, Hungary\\
$^{2}$ MTA-ELTE EIRSA "Lend\"ulet" Astrophysics Research Group, 1117 P\'azm\'any P\'eter s\'et\'any 1/A Budapest, Hungary\\
$^{3}$ Institute for Astronomy, University of Hawaii 2680 Woodlawn Drive, Honolulu, HI, 96822, USA\\
$^{4}$INAF OA Brera, Via E. Bianchi 46, Merate, Italy\\
$^{5}$Department of Physics and Astronomy, The Johns Hopkins University, Baltimore MD 21218, USA\\
$^{6}$Department of Physics, Durham University, South Road, Durham DH1 3LE, UK\\
$^{7}$Department of Astrophysical Sciences, Princeton University, Princeton, NJ 08544
}

\pubyear{2014}
\volume{306}  
\pagerange{119--126}
\jname{Statistical Challenges in 21st Century Cosmology}
\editors{A.C. Editor, B.D. Editor \& C.E. Editor, eds.}
\begin{document}

\maketitle

\begin{abstract}
We use a WISE-2MASS-Pan-STARRS1 galaxy catalog to search for a supervoid in the direction of the Cosmic Microwave Background Cold Spot. We obtain photometric redshifts using our multicolor data set to create a tomographic map of the galaxy distribution. The radial density profile centred on the Cold Spot shows a large low density region, extending over 10's of degrees. Motivated by previous Cosmic Microwave Background results, we test for underdensities within two angular radii, $5^\circ$, and $15^\circ$. Our data, combined with an earlier measurement by \cite{GranettEtal2010}, are consistent with a large $R_{\rm void}=(192 \pm 15)\mpc $ $(2\sigma)$ supervoid with $\delta \simeq -0.13 \pm 0.03$ centered at $z=0.22\pm0.01$. Such a supervoid, constituting a $\sim3.5 \sigma$ fluctuation in the $\Lambda CDM$ model, is a plausible cause for the Cold Spot.
\end{abstract}

\section{Introduction}

The Cosmic Microwave Background (CMB) Cold Spot (CS) is an exceptionally cold area centered on on $(l,b) \simeq (209^\circ,-57^\circ)$ and was first detected in the Wilkinson Microwave Anisotropy Probe (\cite{bennett2012}) at $\simeq 3\sigma$  significance using wavelet filtering (\cite{VielvaEtal2003}). The CS is perhaps the most significant among the ÒanomaliesÓ recently confirmed by {\it Planck} (\cite{Planck23}). These physical anomalies are significant enough to motivate further studies. Some of the models include exotic physics, e.g., cosmic textures (\cite{CruzEtal2008}), while e.g. \cite{Inoue2006} claim that the CS, and possibly other anomalies, are caused by the Integrated Sachs-Wolfe effect (ISW) of the decaying gravitational potentials, which in turn is caused by Dark Energy. The latter explanation would require of a large, $\gtrsim 200\mpc$ supervoid centered on the CS, readily detectable in large scale structure surveys.

The supervoid model has been constrained by using radio galaxies of the NVSS survey, Canada-France-Hawaii Telescope (CFHT) imaging of the CS region, redshift survey data using the VIMOS spectrograph on the VLT, and the relatively shallow 2MASS galaxy catalogue. See \cite{SzapudiEtal2014}, and references therein for review. Although these works report a possible underdense region at redshifts $z < 0.3$, they either run out of objects at low redshift, or have no redshifts for tomographic imaging. Note that although no void has been found that could fully explain the CS anomaly, there is strong, $\gtrsim 4.4\sigma$, statistical evidence that superstructures imprint on the CMB as cold and hot spots (\cite{GranettEtal2008}).

\section{Analysis of WISE-2MASS-PS1 galaxy counts}

Next we describe some of our methods and procedures, while \cite{SzapudiEtal2014} and \cite{FinelliEtal2014} can be consulted for further details.
We probe the $z \simeq 0.3$ redshift range, unconstrained by previous studies, using the WISE-2MASS galaxy catalog (\cite{KovacsSzapudi2014}) with $z_{\rm med}\simeq 0.14$, and 21,200 square degrees of coverage. The catalog contains sources to flux limits of $W1_{WISE}$ $\leq $ 15.2 mag and $J_{2MASS}$  $\leq $ 16.5 mag, resulting in a galaxy sample deeper than 2MASS (\cite{2MASS}) and more uniform than WISE (\cite{wise}). 

In this projected catalog, we detect an underdensity centred on the Cold Spot. In particular, we find signal-to-noise ratios $S/N \sim 12$ for rings at our pre-determined radii, $5^\circ$ and $15^\circ$, and an extended underdensity to $\sim 20^\circ$ with $\gtrsim 5\sigma$ significance. At larger radii, the radial profile is  consistent with a supervoid surrounded by a gentle compensation that converges to the average galaxy density at $\sim 50^\circ$. The supervoid appears to contain an extra depression in its centre, with its own compensation at around $\simeq 8^\circ$. The projected underdensity in WISE-2MASS is modeled with a $\Lambda$LTB model (\cite{GBH2008}) by \cite{FinelliEtal2014}.

We then set out to create a detailed map of the CS region in three dimensions. Therefore, WISE-2MASS galaxies have been matched with Pan-STARRS1 (\cite{ps1ref}, PS1) objects within a $50^\circ\times50^\circ$ area centred on the CS, except for a $\rm{Dec}\geq-28.0$ cut to conform to the PS1 boundary. We used PV1.2 reprocessing  of PS1 in an area of 1,300 square degrees.
For PS1, we required a proper measurement of Kron and PSF magnitudes in $g_{\rm P1}$, $r_{\rm P1}$ and $i_{\rm P1}$ bands that  were used to construct photometric redshifts (photo-$z$'s) with a Support Vector Machine algorithm. The training and control sets were created matching WISE-2MASS, PS1, and the Galaxy and Mass Assembly \cite{gama} (GAMA) redshift survey. Our three dimensional catalog of WISE-2MASS-PS1 galaxies contains photo-$z$'s with an estimated error of $\sigma_{z} \approx 0.034$.
\begin{figure}
\begin{center}
\includegraphics[height=60mm]{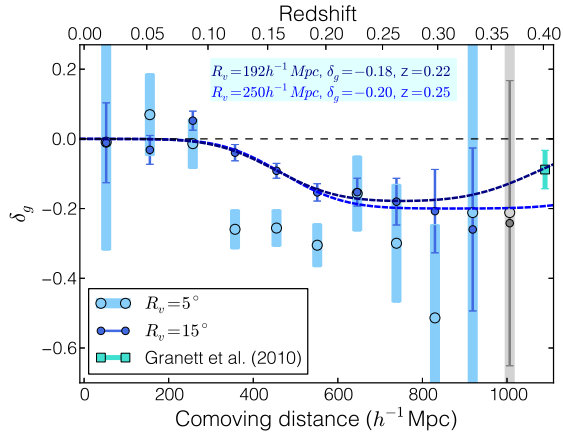}
\caption{Our measurements of the galaxy density in the line-of-sight using the $\Delta z = 0.035$ photo-$z$ bins we defined. We detected a significant depression in galaxy density in $r=5^{\circ}$ and $15^{\circ}$ test circles. We used our simple modeling tool to examine the effects of photo-$z$ errors, and test the consistency of simple top-hat voids with our measurements. The last bin (shown in grey) was excluded from the analysis, and a data point by Granett et al (2010) accounts for the higher redshift part of the measurement. See text for details.}
\label{fig_pz}
\end{center}
\end{figure}
We count galaxies as a function of redshift in disks centred on the CS using the two pre-determined angular radii, $R=5^\circ$, and $15^\circ$. The galaxy density calculated from the average redshift distribution is shown in the upper panel of Fig.~\ref{fig_pz}. Note that the larger disk is not fully contained in our photo-$z$ catalog due to the limited PS1 footprint.
In photo-$z$ bins of width of $\Delta z = 0.035$, we found $S/N \sim 5$ and $S/N \sim 6$ for the deepest under-density bins, respectively. The measurement errors are due to Poisson fluctuations in the redshift bins. To extend our analysis to higher redshifts, we add a previous measurement (\cite{GranettEtal2010}) in a photo-$z$ bin centred at $z=0.4$.

For a rudimentary understanding of our counts, we build toy models from top-hat voids in the $z$ direction convolved with the photo-$z$ errors. This model has  three parameters, redshift ($z_{\rm void}$), radius ($R_{\rm void}$), and depth ($\delta_{g}$).
We carry out the $\chi^{2}$-based maximum likelihood parameter estimation for $R=15^{\circ}$, using our first 10 bins combined with the extra bin measured by \cite{GranettEtal2010} for $n=11$ data points.   We find $\chi^{2}_{0} = 92.4$ for the null hypothesis of no void. The marginalized results for the parameters of our toy model are $z_{\rm void}=0.22\pm 0.01$ $(2\sigma)$,  $R_{\rm void}=(192 \pm 15)\mpc $ $(2\sigma)$, and $\delta_{g}=-0.18\pm 0.01$ $(2\sigma)$ finding $\chi^{2}_{min} = 5.97$. We take the $b_{g} = 1.41 \pm 0.07$ galaxy bias of the catalog into account finding an average depth of $\delta = \delta_{g} / b_{g} \simeq -0.13 \pm 0.03$ ($2\sigma$).
\begin{figure}
\begin{center}
\includegraphics[width=145mm]{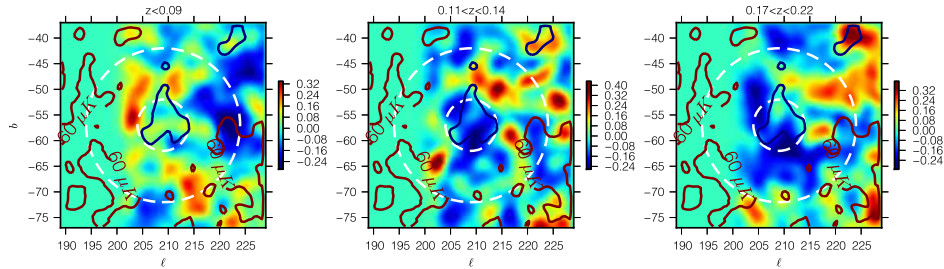}
\caption{Tomographic view of the CS region. Below $z < 0.09$ appears to show the front compensation (higher density area) of the large void, although there is an under dense structure to the right. The slice at $0.11 < z < 0.14$  cuts into front of the supervoid, and the ring round represents a slice of the compensation. Finally the slice $0.17 < z < 0.22$ cuts into the front half of the supervoid with compensation around it. Note that the left side of the images reflects the mask of the PS1 data set.}
\label{fig_visu}
\end{center}
\end{figure}
While the toy model quantifies the properties of the void, careful inspection of the Figures reveals a complex structure. There appears to be a compensation in front of the supervoid at around $z\simeq 0.05-0.08$, and the significantly deeper counts at the smaller radii show that the void is deeper at the centre. An approximate tomographic imaging of the CS region is presented in Fig.~\ref{fig_visu}. in three slices of $z < 0.09$, $0.11 < z < 0.14$,  $0.17 < z < 0.22$ and smoothed at $2^{\circ}$ scales. The Planck SMICA CMB temperature map is over-plot with contours. While we characterized the profile in radial bins, the geometry of the supervoid is more complex. It is noteworthy that deepest part of the void is close to the centre of the CS in the middle slice. Given the enormous size, the rich structure is expected: the supervoid contains voids, their compensations, filaments, and appears to have its own compensation.

\section{Conclusions}

We have found strong statistical evidences for a supervoid at redshift $z = 0.22\pm 0.01$ with radius $R = (192\pm 15)\,h^{-1}$Mpc and depth of $\delta = -0.13 \pm 0.03$ (\cite{SzapudiEtal2014}). These parameters are in excellent agreement with the projected density analysis results by \cite{FinelliEtal2014}, with the redshift slightly closer.
We estimated the linear ISW effect due to such an underdensity from a simple approximation. We found that it could significantly affect the CMB,  of order $-20$ $\mu$K, falling short of a full explanation of the CS anomaly. However, a non-linear LTB supervoid based on the projected profile in the WISE-2MASS catalog matches well the profile observed on the CMB (\cite{FinelliEtal2014}).
The observed alignment and morphological similarity to a compensation is noteworthy. We estimate that the supervoid we detected corresponds to a rare, at least $\gtrsim 3.5\sigma$, density fluctuation in \LCDM; thus chance alignment with another rare structure, the CS, is negligible.

\section*{Acknowledgments}
\small{
We acknowledge the support of NASA grants NNX12AF83G and NNX10AD53G. AK and ZF acknowledge support from OTKA through grant no. 101666. In addition, AK acknowledges support from the Campus Hungary fellowship program. We make use of HEALPix \cite{healpix} software in our project.
The Pan-STARRS1 Surveys (PS1) have been made possible through contributions by the Institute for Astronomy, the University of Hawaii, the Pan-STARRS Project Office, the Max-Planck Society and its participating institutes, the Max Planck Institute for Astronomy, Heidelberg and the Max Planck Institute for Extraterrestrial Physics, Garching, The Johns Hopkins University, Durham University, the University of Edinburgh, the Queen's University Belfast, the Harvard-Smithsonian Center for Astrophysics, the Las Cumbres Observatory Global Telescope Network Incorporated, the National Central University of Taiwan, the Space Telescope Science Institute, and the National Aeronautics and Space Administration under Grant No. NNX08AR22G issued through the Planetary Science Division of the NASA Science Mission Directorate, the National Science Foundation Grant No. AST-1238877, and the University of Maryland, and Eotvos Lorand University (ELTE).
}


\begin{thebibliography}{}

\bibitem[Inoue and Silk 2007]{Inoue2006}
{Inoue, K.~T. \& Silk, J.} 2006,
\textit{ApJ}, 648, 23-30

\bibitem[Szapudi et al. 2014]{SzapudiEtal2014}
{Szapudi, I. and Kov{\'a}cs, A., Granett, B.~R., et al.} 2014,
\textit{arXiv:1405.1566}

\bibitem[Finelli et al. 2014]{FinelliEtal2014}
{Finelli, F.,  Garcia-Bellido, J., Kovacs, A., Paci, F., and Szapudi, I.} 2014,
\textit{arXiv:1405.1555}

\bibitem[Kov{\'a}cs and Szapudi 2014]{KovacsSzapudi2014}
{Kov{\'a}cs, A. and Szapudi, I.} 2014,
\textit{arXiv:1401.0156}

\bibitem[Garcia-Bellido and Haugb{\o}lle 2008]{GBH2008}
{Garcia-Bellido, J. and Haugb{\o}lle, T.} 2008,
\textit{JCAP}, 4:3

\bibitem[Gorski et al. 2005]{healpix}
{K.~M. Gorski, E.~Hivon and et~al.} 2005,
\textit{Apj}, 622:759--771

\bibitem[Wright et al. 2010]{wise} {E.~L. Wright et~al.} 2010,
\textit{AJ}, 140:1868--1881

\bibitem[Granett et al. 2008]{GranettEtal2008}
{B.~R. Granett, M.~C. Neyrinck, and I.~Szapudi} 2008,
\textit{ApjL}, 683:L99--L102

\bibitem[Granett et al. 2010]{GranettEtal2010}
{B.~R. {Granett}, I.~{Szapudi}, and M.~C. {Neyrinck}} 2010,
\textit{Apj}, 714:825--833

\bibitem[Skrutskie et al. 2006]{2MASS}
{M.~F. {Skrutskie}, R.~M. {Cutri}, R.~{Stiening}, and et~al.} 2006,
\textit{AJ}, 131:1163--1183

\bibitem[Bennett et al. 2012]{bennett2012}
{C.~L. {Bennett}, D.~{Larson}, and J.~L et~al. {Weiland}.} 2012
\textit{ArXiv e-prints}

\bibitem[Vielva et al. 2003]{VielvaEtal2003}
{P.~{Vielva} et al.} 2004,
\textit{Apj}, 609:22--34.

\bibitem[Cruz et al. 2008]{CruzEtal2008}
{M.~{Cruz} et al.} 2008,
\textit{MNRAS}, 390:913--919

\bibitem[Driver et al. 2011]{gama}
{S.~P. {Driver}, D.~T. {Hill}, and et~al.} 2011
\textit{MNRAS}, 413:971--995

\bibitem[Planck 2013 results. XXIII.]{Planck23}
{Planck 2013 results. XXIII.} 2013
\textit {ArXiv e-prints}

\bibitem[Kaiser 2004]{ps1ref}
{N.~{Kaiser}.}
\textit{SPIE Conference Series}, 2004.

\end{thebibliography}
\end{document}